\documentclass[tradiabstract]{aa}
\usepackage{graphicx}
\usepackage{longtable}
\usepackage{latexsym}
\usepackage{amssymb}
\usepackage{lscape}
\usepackage{natbib}
\usepackage{txfonts}

\begin{document}

\title{HI content and other structural properties of galaxies in the Virgo cluster from the Arecibo Legacy Fast ALFA Survey.}

\author{Giuseppe Gavazzi\inst{1}, 
\and Riccardo Giovanelli\inst{2},   
\and Martha P. Haynes\inst{2},
\and Silvia Fabello\inst{1}, Michele Fumagalli\inst{1},
\and Brian R. Kent\inst{2},
\and Rebecca A. Koopmann\inst{3}, 
\and Noah Brosch\inst{4},
\and G. Lyle Hoffman\inst{5},
\and John J.  Salzer\inst{6},
\and Alessandro Boselli\inst{7} 
}

\institute{Universita degli Studi di Milano-Bicocca, Piazza delle Scienze 3, 20126
Milano, Italy
\and Center for Radiophysics and Space Research and National Astronomy and Ionosphere Center, Cornell
University, Ithaca, NY 14853
\and Dept. of Physics \& Astronomy. Union College. Schenectady, NY 12308
\and The Wise Observatory and The Raymond \& Beverly Sackler School of Physics and Astronomy, Tel Aviv University, Israel
\and Hugel Science Center, Lafayette College, Easton, PA 18042
\and Astronomy Dept. Wesleyan University, Middletown, CT 06457
\and Laboratoire d'Astrophysique de Marseille, BP8, Traverse du Siphon, F-13376 Marseille, France}

\date{Accepted 25/01/2008}

\abstract{We report the results of an HI blind survey of $\rm 80 ~deg^2$ of the Virgo cluster, based on the $08^o \leq \delta \leq 16^o$ strip of
ALFALFA, the Arecibo Legacy Fast ALFA Survey.
187 HI sources of high significance are found providing a complete census of 
HI sources in this region of the Virgo cluster ($-1000<cz<3000$ $\rm km~s^{-1}$) with $M_{HI} \geq 10^{7.5 \sim 8} ~M_\odot$.
156/187 (83\%) sources are identified with optical galaxies from the Virgo Cluster Catalogue
(Binggeli et al. 1985), all but 8 with late-type galaxies. 
Ten sources are not associated with optical galaxies and were found
to correspond to tidally-disrupted systems (see Kent et al. 2007 and Haynes, Giovanelli and Kent 2007). 
The remaining 21 (11\%) 
are associated with galaxies that are not listed in the Virgo Cluster Catalogue. 
For all sources with an optical counterpart in the Sloan Digital Sky Survey, we analyzed $i$-band SDSS plates
to measure optical structural parameters. 
We find that in the Virgo cluster: 
i) HI inhabits galaxies that are structurally similar to ordinary late-type galaxies;  
ii) their HI content can be predicted from their optical luminosity;
iii) low surface brightness galaxies have low optical luminosity and contain small
quantities of neutral hydrogen; 
iv) low surface brightness, massive Malin1 type galaxies are comfortably rare objects (less than 0.5 \%);
v) there are no ``dark-galaxies" with HI masses $M_{HI} \geq 10^{7.5 \sim 8} ~M_\odot$;  
vi) less than 1\% of early-type galaxies 
contain neutral hydrogen with $M_{HI} \geq 10^{7.5 \sim 8} ~M_\odot$ (di Serego Alighieri et al. 2007).}

\titlerunning{Properties of HI selected galaxies in the Virgo cluster}
\authorrunning{G. Gavazzi et al.}

\maketitle

\section{Introduction}

The  faint-end slope of the halo mass function either predicted 
analytically (Press \& Schechter 1974) ($\alpha=-1.8$) or by numerical CDM simulations 
(e.g. Jenkins et al. 2001)($\alpha=-2$)
is uncomfortably steeper than the observed slope of the optical luminosity 
function of galaxies (e.g. Blanton et al. 2003) ($\alpha=-1.1$) (see however $\alpha=-1.5$ in $r$ band
by Blanton et al. 2005).
If many sterile halos (i.e. unable to give birth to stars), the so called 
$dark~galaxies$, or many low mass galaxies with low optical surface brightness, yet retaining
some neutral hydrogen existed, outnumbering the normal galaxies,
the faint-end slope of the HI mass function would be steeper than that of the
optical luminosity function, perhaps reconciling the observations with the theoretical predictions. 
However the measured faint-end slope of the HI mass function, even in its most robust determination
(Springob et al. 2005,) is significantly flatter ($\alpha=-1.24$) 
than the theoretical one (see also Zwaan, Briggs \& Sprayberry, 2001). 
The Zwaan et al. (2005) HI mass function,
based on HIPASS (HI Parkes All-Sky Survey) data (Meyer et al. 2004), 
is flatter ($\alpha=-1.37$) than predicted, although it
suffers from  statistical limitations due to the paucity of objects (40) 
with $M_{HI} \leq 10^8 M_\odot$ and from distance uncertainties (Masters et al. 2004).
To improve the determination of the HI mass function, several blind HI surveys have been recently
carried out or are under way. 
HIDEEP (Minchin et al. 2003) is a deep HI survey of $\rm 32~deg^2$ in Centaurus, carried out with the 
Parkes multibeam system, sensitive to $M_{HI} \geq  10^{8} M_\odot$, with deep optical follow-up plates.
A more recent one that combines a large area with a high sensitivity
is the $\rm 86~deg^2$ survey in the direction
of Canes Venatici (Cvn) carried out with the WSRT (Westerbork Synthesis Radio Telescope) by Kovac (2007). 
Incidentally the slope of the faint-end 
mass function derived using
70 sources with $M_{HI} \geq  10^{6.5} M_\odot$ from this survey
is flatter than ever ($\alpha=-1.17$).\\
The Arecibo telescope, equipped with the new 7 beam system ALFA, has recently
started a number of HI blind surveys.
The Arecibo Galaxy Environment survey (AGES) (Auld et al. 2006) has covered some patches of sky, including $\rm 5~deg^2$ of the Coma supercluster containing the
cluster A1367, with a sensitivity of $M_{HI} =  10^{8.8} M_\odot$ at the distance of Coma 
(300 s integration time per beam, rms=0.84 mJy/beam; Cortese et al. 2007). 
Highly statistically significant results will be obtained from ALFALFA, 
the Arecibo Legacy Fast ALFA Survey, that covers the Arecibo sky ($\rm 7000 ~deg^2$) in a 2-pass drift-scan mode
(48 s integration, with a typical rms=2.1 mJy/beam at 10 km/s resolution; Giovanelli et al. 2005).
Besides a robust determination of the faint-end slope of the HI mass function, 
(based on several hundred galaxies with $M_{HI} \leq  10^{7.5} M_\odot$)
ALFALFA will provide more evidence on whether low surface brightness galaxies,
that are so elusive optically (Disney 1976; Disney \& Phillipps 1987; Sabatini et al. 2003),
will change our understanding of galaxies in the local universe. 
This issue has two aspects that could be addressed and solved with ALFALFA.
One is to increase the number of dwarf galaxies surveyed to better determine the number density of low surface brightness
galaxies that, in spite of their low optical visibility, are HI rich. 
(Galaxies of decreasing mass have an increasing gas fraction 
and a decreasing optical ($i$-band) surface
brightness (Boselli et al. 2001; Warren et al. 2006)).
The other is to confirm and corroborate the finding of 
Briggs (1997), Rosenberg et al (2002), Zwaan et al (2003),
Minchin et al. (2004) that the frequency of Malin1-type galaxies, i.e.
massive, gas rich objects of very low optical surface brightness, is insignificant (see also Hayward, Irwin \& Bregman 2005).
Malin1 was indeed discovered in HI in the background of the Virgo cluster (Impey \& Bothun 1989).  
How many unknown low surface brightness galaxies remain hidden in the local universe, 
unseen by optical observations? ALFALFA should provide the complete census of such objects.\\
In the last two years ALFALFA has covered most of the Virgo cluster (Giovanelli et al. 2006), the nearest rich cluster,
providing $\sim$ 1000 HI sources in $\rm 80~deg^2$, 
perhaps the highest space density of HI sources in the whole ALFALFA survey.
The analysis of this sample will dramatically enhance our understanding of the issues outlined above.
The Virgo region is especially useful because of
the existence of a very deep optical catalogue of galaxies, the Virgo Cluster Catalogue (VCC) 
by Binggeli et al. (1985), that will allow a comparison of the properties of optically and radio selected
objects in this region. 
The overlap between the ALFALFA survey and Goldmine, a collection 
of HI pointed observations in the direction of Virgo obtained by many observers during the years (Gavazzi et al. 2005),
also allows a comparison of the relative quality of the two databases.
These are the issues that the present paper on the Virgo cluster, as seen by ALFALFA, wishes to address.
A forthcoming paper will analyze $H\alpha$ observations obtained recently of several hundred HI galaxies discovered by ALFALFA
in the Virgo cluster and in its immediate surroundings
(Gavazzi et al. in preparation), addressing
the issue of the environmental dependence of galaxy evolution (see a review by Boselli \& Gavazzi 2006) by comparing
the rate of transformation of primordial gas into young stars in the Virgo
cluster with that in less dense environments.\\
The structure of the present paper is as follows. 
The HI and optical selected samples are 
presented in Section 2. The data analysis is presented in Section 3, where
the reduction  of optical images taken from the Sloan Digital Sky Survey (SDSS) is discussed in detail. 
Section 4 gives the comparison between the results of ALFALFA and pointed observations available
in the literature. The results are discussed in Section 5. 
  
\section{The sample}

The present analysis is focused on the ALFALFA blind HI survey in the region of the Virgo cluster;
more precisely in the area of intersection between ALFALFA, as available in Sept 2007,
and the VCC.
The ALFALFA survey provides a sample of HI selected objects, and the VCC a list of
optically selected galaxies. 

\subsection{The HI selected sample}

The HI selected sample consists of 187 ALFALFA sources meeting the following
criteria:\\
i) their right ascensions lie in the interval $12^h08^m30^s \leq R.A. \leq 12^h48^m20^s$ 
(ca. $182^o\leq R.A. \leq 192^o$) (J2000) corresponding to the extent in R.A. of 
the VCC, and their declinations range between
$08^o \leq \delta \leq 16^o$ (J2000), i.e., they lie in the declination strip that has been
fully mapped by ALFALFA. North of 12 deg the catalogued sources are taken from 
Giovanelli et al. (2007); for $08^o \leq \delta \leq 12^o$ the list of sources 
is not yet published (Kent et al. in preparation);\\
ii) they have recessional velocities in the range $-1000<cz<3000$ $\rm km~s^{-1}$,
bracketing the full depth of the Virgo cluster (see Gavazzi et al. 1999);\\
iii) they have quality Code = 1, i.e., their statistical significance is higher than S/N=6.5.
(We do not consider Code = 2 objects, i.e., sources with S/N $<$ 6.5 that have 
optical counterparts
with optical redshifts matching the HI line redshift.)\\
iv) they are not included in the circle of projected radius of 1.0 deg centered
on M87, where ALFALFA is strongly incomplete because of loss of spectral sensitivity due to the
bright (220 Jy at 1415 MHz) continuum source associated with M87;\\
v) they lie outside one resolution element (3.5 arcmin) of any other 
strong source at similar redshift. The six excluded sources because of this criterion
are:\\
122140.1 +143621 = VCC0497 conf by VCC0483\\ 
122140.1 +143621 = VCC1673 conf by VCC1676\\
124331.5 +113500 = VCC1972 conf by VCC1978\\

Since the mean rms of ALFALFA is 2.1 mJy/beam, 
the survey detection limit (for S/N=6.5) is 0.5 $\rm Jy/beam ~km ~s^{-1}$ for typical widths of 40 $\rm km ~s^{-1}$, 
thus, at the mean
distance of the Virgo cluster (17 Mpc), using the relation:
\begin{equation}
M_{HI} = S/N \times rms \times W \times dist^2 \times 2.36 \times 10^5 ~(M_\odot)
\end{equation}
these sources have $M_{HI}\geq10^{7.5} ~M_\odot$. 
The $M_{HI}$ limit, however depends on the line width, thus on the system's mass and inclination, as detailed in Section 5.1,
resulting in $10^{7.5}~M_\odot  \leq M_{HI~limit} \leq 10^8 ~M_\odot$.

\subsection{The optically selected sample}

From the VCC catalogue (Goldmine version, Gavazzi et al. 2003), restricted to
the region in common with ALFALFA 
we have extracted two subsamples of optically selected galaxies that are 
bona-fide Virgo cluster members, whose distances have been assigned following the
subcluster membership criteria of Gavazzi et al. (1999), updated with new redshift measurements.\\
The VCC catalogue contains galaxies as faint as 20 mag, but is 
complete to $m_B<18$ mag, within a limiting surface brightness of
25.3 B $\rm mag~arcsec^{-2}$ (Binggeli et  al. 1985). From this catalogue we
have considered:\\ 
Sample A (deep) consisting of 1112 galaxies with $m_B<20$ mag;\\
Sample B (shallow) consisting of 469 galaxies with $m_B \leq 17$ mag\footnote{We 
have preferred not to extend the shallow sample B
to $m_B = 18$ mag because in the bin  $17\leq m_B \leq 18$ mag only 6\% of the VCC galaxies have an HI counterpart
in ALFALFA, while in the bin $16\leq m_B \leq 17$ mag 40\% are in ALFALFA.}, 
278 of which are early-type (dE-E-S0-S0a)\footnote{Notice that the morphological class ``dE" might be contaminated 
by some transition objects (dE/dIrr) but we find at most 8 gas-rich ones, as shown in Section. 3.4.}  and 191 are late-type (Sa-Irr-BCD) galaxies.
The number of galaxies with redshifts is 410/469, including
226/278 early-types and 184/191 late-types.
At the mean distance of the Virgo cluster (17 Mpc, or $M-m=-31.1$ mag) $m_B \leq 17$ mag corresponds to $M_B \leq -14.1$ mag,
thus even Sample B includes dwarf systems. 

\section{The data reduction}

\subsection{HI (ALFALFA)}

\begin{figure}
\includegraphics[width=9cm,angle=0]{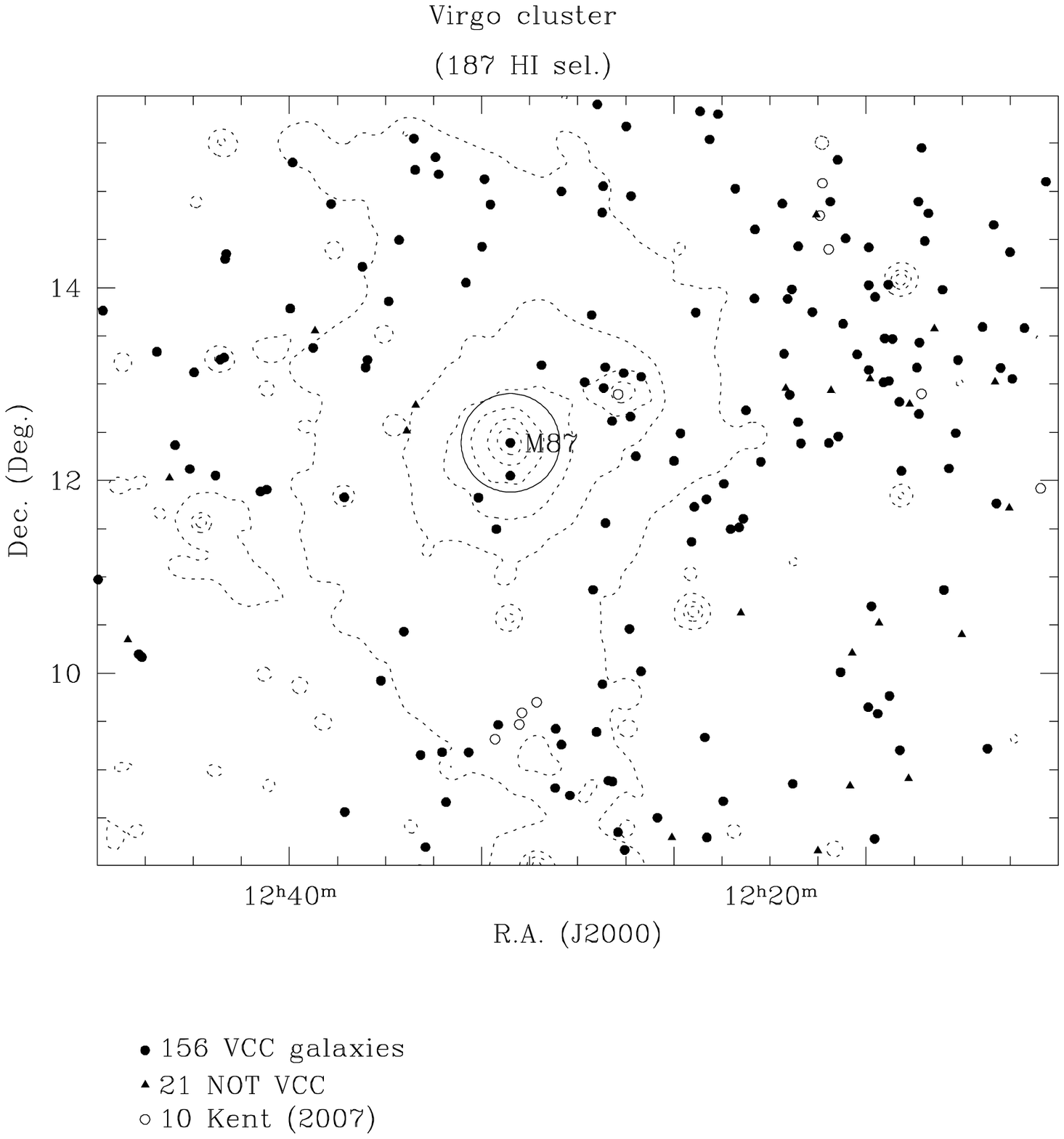}
\includegraphics[width=9cm,angle=0]{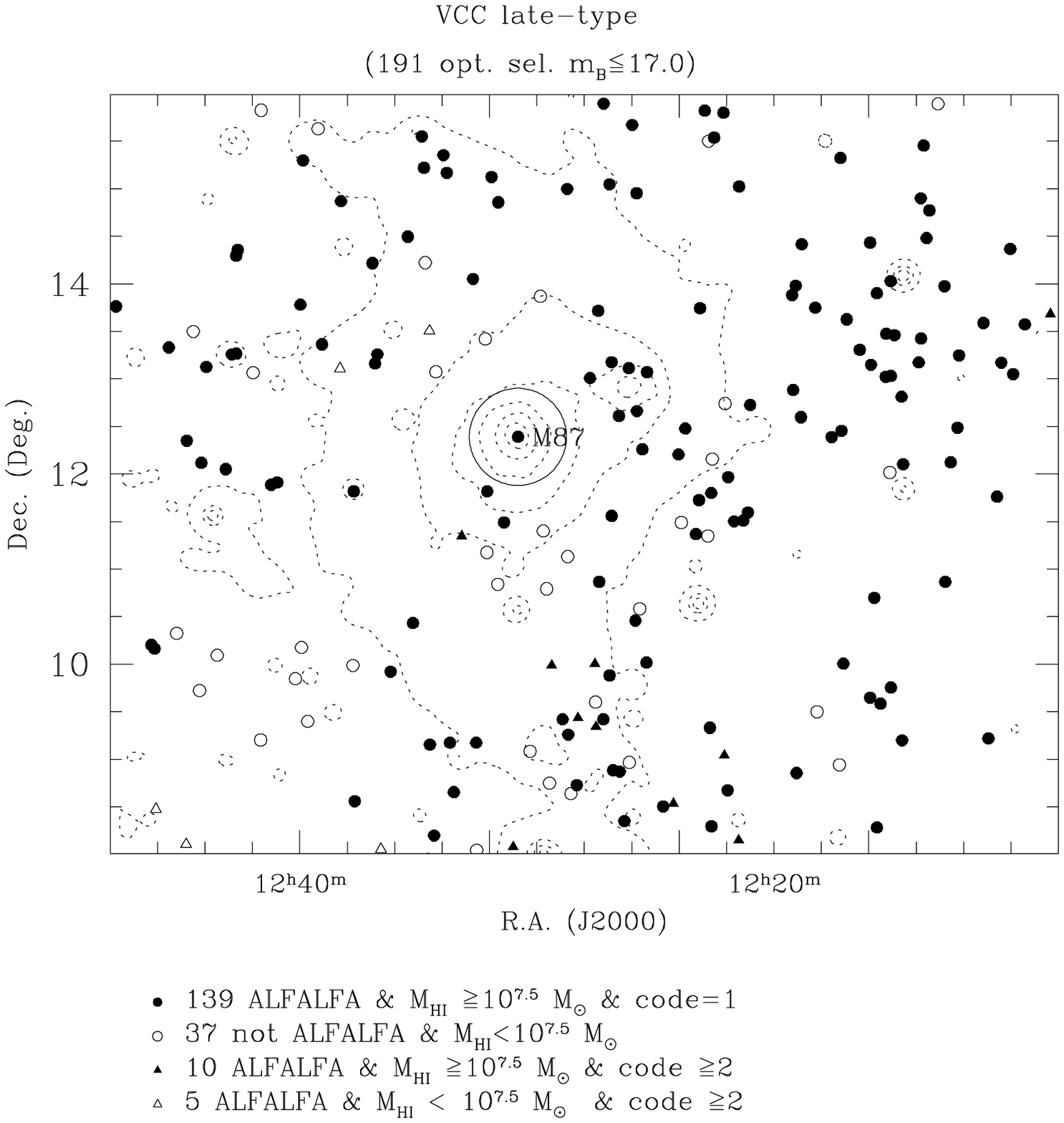}
\caption{The distribution in celestial coordinates of sources in the Virgo cluster. Contours
from the X-ray ROSAT all sky survey (B\"ohringer et al. 1994) are given by dashed lines. 
The solid line shows a one degree circle centered on M87, within which ALFALFA
is complete due to the strong continuum of M87.
Top: ALFALFA selected sources identified with VCC galaxies with $m_B<20$ (filled circles). 
Ten non-identified 
sources labeled as tidally-disrupted clouds by Kent et al. (2007) (empty circles) and 21 sources identified
with faint galaxies not listed in the VCC (filled triangles) are given.
Bottom: Optically selected galaxies, consisting of 191 late-type galaxies with $m_B \leq 17$ in the VCC. 
139 that have been detected by ALFALFA are plotted with filled circles, 15 that have been detected by
ALFALFA with Code $\geq 2$ with filled triangles and another 37 detected in Goldmine, but not
ALFALFA (as their HI mass is $M_{HI} <10^{7.5} ~M_\odot$ based on Goldmine) with
empty circles.}
\label{figure1}
\end{figure}
\begin{figure}
\includegraphics[width=9cm,angle=0]{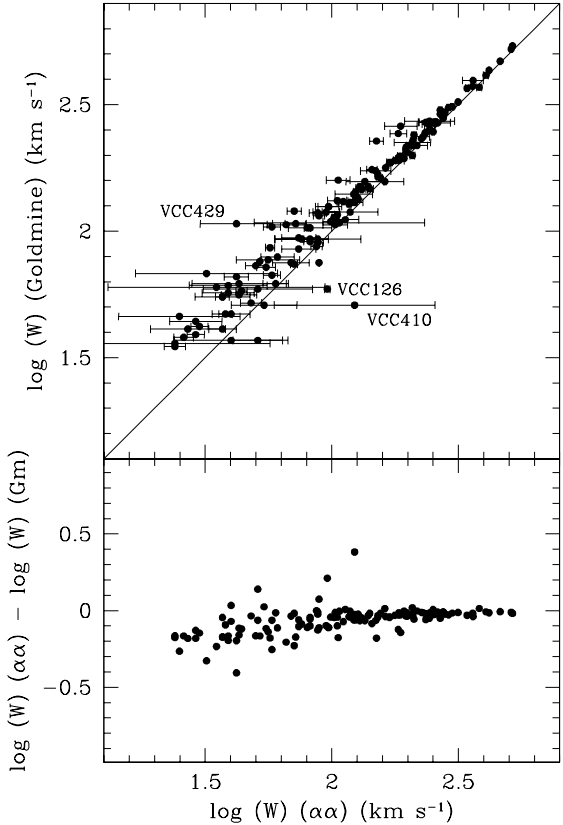}
\small{\caption{Comparison of the ALFALFA (blind) and Goldmine (pointed) 21-cm line width measurements.
The most deviant objects are labeled with their VCC names.
The line represents the one to one relation.}
\label{figure2}}
\end{figure}
For each of the HI selected sources we consider the following HI parameters taken from ALFALFA
(Giovanelli et al. 2007):\\
- the (J2000) coordinates (affected by an error that depends on S/N, with a mean of $\approx$ 24 arcsec); \\
- the recessional velocity $VP$; \\
- the observed width of the HI line $W$. This is the width of the profile measured at
the 50\% level of each of the two peaks, corrected for instrumental broadening; \\
- the total flux under the HI line: $SintP$ ($\rm Jy~km~s^{-1}$);\\
For sources optically identified with VCC galaxies, we convert $SintP$ into the HI mass using:
$M_{HI}=2.36 \times 10^5 \times SintP \times (dist)^2 ~(M_\odot)$, 
where the distance is determined as in Gavazzi et al. (1999), otherwise using 17 Mpc, and
we correct the observed line width $W$ for projection effects, using
the ratio of the optical minor-to-major axes:
$W_c= W/sin(acos(b/a))$.\\
We also estimate the HI deficiency parameter following Haynes \& Giovanelli (1984)
as the logarithmic difference between $M_{HI}$ of a reference sample of isolated 
galaxies and $M_{HI}$ actually observed in individual objects: $Def_{HI}= Log M_{HI~ref.} - Log M_{HI~obs.}$. 
$Log M_{HI~ref}$ has been found to be linearly related to the galaxies linear diameter $d$ as: 
$Log M_{HI~ref}=a+b Log(d)$, where $d$ (in kpc) is determined at the $25^{th}$ B-band isophote,    
and $a$ and $b$ are weak functions of the Hubble type. 
We caution that the Haynes \& Giovanelli (1984) 
reference sample of isolated galaxies included only relatively large ($a>1$ arcmin) UGC objects 
so that the $Def_{HI}$ parameter is poorly calibrated for smaller objects,
making determinations of the HI deficiency for the smallest objects uncertain and
likely underestimated (Solanes 1996). 
Furthermore, as discussed in Solanes et al. (2001) galaxies in the latest
Hubble types (Scd-Im-BCD), for which we have adopted $a$ and $b$ parameters
consistent with those of Sc, are more subject to observational biases
than higher surface brightness galaxies. The reader should be aware that
the determinations of the HI deficiency for these objects is highly uncertain. \\
For all radio sources we compute the total flux divided
by the line width: $\Sigma_{HIc}=SintP/W$ in Jy/beam; 
this parameter is an indicator of the "visibility" in the ALFALFA survey (see Fig.\ref{figure8}).\\

\subsection{Optical \& HI (Goldmine)}

\begin{figure}[t]
\includegraphics[width=9cm,angle=0]{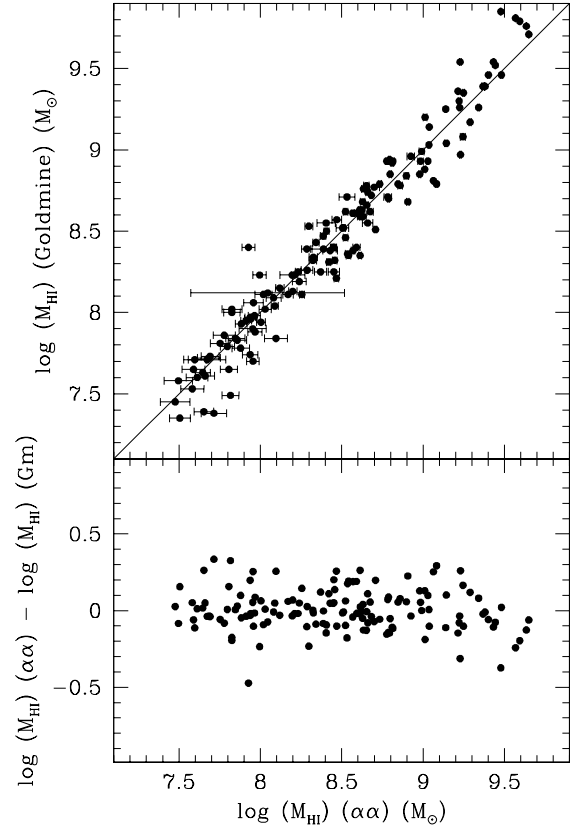}
\small{\caption{Comparison of the ALFALFA (blind) and Goldmine (pointed) 
21-cm line intensity measurements of Virgo galaxies. The line represents the one to one relation.}
\label{figure3}}
\end{figure}

For each of the VCC galaxies we take from the Goldmine database:\\
- the (J2000) optical coordinates (accurate to $\approx$ 30 arcsec); \\
- the morphological type (adopted from Binggeli et al. 1985 and from Binggeli et al. 1993);\\
- the major ($a$) and minor ($b$) axes (arcmin) (from Binggeli et al. 1985). These
quantities are used to correct the observed line widths according to the galaxy inclination;\\
- the distance (in Mpc) (computed according to the criteria of Gavazzi et al. 1999).
The adopted distances for each subgroup of Virgo are: $17\pm 0.3$ Mpc  for cluster A, 
the North and East clouds and the Southern extension, $23\pm 0.5$ Mpc
for cluster B and $32\pm 0.9$ Mpc for the W and M clouds.\\
In order to compare the new HI parameters derived by ALFALFA with 
the plethora of earlier pointed observations, collected and made as 
homogeneous as possible by Gavazzi et al. (2005)for the Goldmine HI database, 
we also consider:\\
- the measured width of the HI line. 
The Goldmine measurement differs from that
of ALFALFA, as it corresponds to the average of the line widths measured
at the 20\% and 50\% levels of the peak intensity, 
uncorrected for instrumental broadening;\\
- the hydrogen mass $M_{HI}=2.36 \times 10^5 \times Sint \times (dist)^2 ~(M_\odot)$ 
where $Sint$ is derived from pointed 21 cm observations.
It should be noted that the 21 cm data in Goldmine were obtained from a variety of observations taken at Arecibo,
some of which lasted much longer (typically one hour) than the ALFALFA observations ($\sim$ 48 sec in the beam),
thus reaching a sensitivity about one order of magnitude better than ALFALFA.
These measurements are however limited
to late-type galaxies. 

\subsection{SDSS imaging material and reduction}

For the optical comparisons, we obtained $g$ and $i$ band images 
from the Sloan Digital Sky Survey (DR 5) (Adelman et al. 2007) that cover
the 187 HI selected sources and the optically selected late-type galaxies with $m_B \leq 17$
(Sample B).
Only one galaxy (VCC1401) was not available the SDSS.
Two or more (up to 4) SDSS fields were combined 
for 13 galaxies (VCC66, 89, 167, 596, 865, 873, 939, 1588, 1727, 1778, 1932, 2058 and 2070)  
that were present in two or more SDSS images.
All frames were photometrically calibrated using the wavelength dependent parameters contained
in the calibration tables associated with the individual SDSS images, namely
the zero point ($aa$), the extinction coefficient ($kk$) and the airmass.
The effective zero point (in AB magnitudes) was derived using:
$ZP_{eff}=aa+kk*airmass$. (See the SDSS data release 6 for further details on the flux calibration.)\\
The SDSS image analysis was carried out 
in the IRAF environment and relied on the STSDAS package
\footnote{IRAF is the Image Analysis and Reduction Facility made
available to the astronomical community by the National Optical
Astronomy Observatories, which are operated by AURA, Inc., under
contract with the U.S. National Science Foundation. STSDAS is
distributed by the Space Telescope Science Institute, which is
operated by the Association of Universities for Research in Astronomy
(AURA), Inc., under NASA contract NAS 5--26555.}  and on GALPHOT
(developed for IRAF- STSDAS mainly by W. Freudling, J. Salzer, and
M.P. Haynes and adapted by S.Zibetti and L. Cortese. See Gavazzi et al. 2001).\\ 
For each frame the sky background was determined 
as the mean number of
counts measured in regions of ``empty'' sky, and it was subtracted
from the frame.  Sky-subtracted frames were inspected individually and
the light from superposed or nearby stars and galaxies was
masked.\\ 
The 2-dimensional light distribution of each galaxy was fit with
elliptical isophotes, using the task $cphot$, a modified version of the STSDAS ${\it
isophote}$ package.  Starting from a set of initial parameters given
manually, the fit maintains as free parameters the ellipse center,
ellipticity and position angle. The ellipse semi-major axis is
incremented by a fixed fraction of its value at each step of the
fitting procedure.  The routine halts when the surface brightness
found in a given annulus equals the sky rms. The fit fails to converge
for 46 galaxies with very irregular light distributions or low surface brightness.  
In these cases we kept fixed one or more of the initial parameters.
In 14 instances of exceedingly low brightness no reliable fit was obtained
(for ten HI selected objects and for four optically selected galaxies, 
namely: VCC 585, 1121, 1582, 1884).\\ The
resulting radial light profiles were fitted with models of the galaxy
light distribution. The majority, 72\%, were best fit with an exponential disk law, while
27\% were best fit by a combination of a disk and a bulge $r^{1/4}$ law and 1\% by
a pure de Vaucouleurs law (de Vaucouleurs 1948).\\
Total magnitudes $g,i$ were then obtained by adding to the flux
measured within the outermost significant isophote the flux
extrapolated to infinity along either 
the exponential law that fitted the outer parts of most galaxies (pure
disks and B+D galaxies), or the $r^{1/4}$ law (dV galaxies).  
The mean statistical uncertainty in the determination of the total magnitude is 0.10 mag.
The effective radius $r_e$ (the radius containing half of the total
light, in arcsec) and the effective surface brightness $\mu_e$ (the mean surface
brightness within $r_e$ in $\rm mag~arcsec^{-2}$) of each galaxy are computed only for the $i$ band.
The mean statistical uncertainties of the determination of $r_e$ and $\mu_e$ are 2 arcsec and
0.15 $\rm mag~arcsec^{-2}$, respectively.
 The statistical uncertanties refer to the errors obtained from fitting models to the observed 
galaxy light distributions; 
we note that they do not correlate with magnitude or surface brightness, but rather with the presence
of complex structures (e.g. bars or irregularities in the light profiles).
Besides the statistical uncertainties, however, the photometric parameters of faint galaxies are affected by
additional uncertainties that might depend on local fluctuations of the sky brightness.
In the following analysis all fits are performed with the maximum likelihood 
method using  statistical uncertainties.\\
Finally we computed another structural parameter: the concentration index
($C_{31}$), defined by de Vaucouleurs (1977) as the model--independent
ratio between the radii that enclose 75\% and 25\% of the total $i$ light.\\
We converted the total $i$-band  magnitude to the stellar mass using:\\
$log (M_{star}/M_\odot)=-0.152+0.518*(g-i)+log I$ (see Appendix A2 in Bell et al. 2003),
where $I$ is the $i$-band luminosity.
The typical uncertainty of $log (M_{star}/I)$ is $\sim 0.1\:dex$.

\subsection{The radio-optical identifications}

We cross-correlated the 187 HI selected sources with the
optically selected sample A (deep) to obtain 
the optical identifications of all HI sources matching a simple
positional criterion (separation$<$1 arcmin). Fig. \ref{figure1}, top panel,
provides the locations of these objects.
We find 156/187 (83\%) identifications with VCC galaxies 
(including 8 sources associated with early type galaxies -
these coincide with the Code 1 sources found by di Serego Alighieri et al. 2007).
Ten of the remaining 31 sources are not associated with optical galaxies and were found
to correspond to tidally-disrupted systems by Kent et al. (2007) and by 
Haynes, Giovanelli and Kent (2007). Among the other  21 (11\% of the total sample), 15  
are associated with faint galaxies that are surprisingly not listed in the VCC in spite of being clearly detected
 and measurable on the SDSS material. Only 6 are not visible in the SDSS images;
 we consider their effective surface brightness fainter than $25 \rm ~mag~arcsec^{-2}$.\\
We then cross-correlated the 469 galaxies in the  
optically selected VCC sample B (shallow) with the ALFALFA positions, separately
for the 191 late-type galaxies and for the 278 early-type galaxies. 
Fig. \ref{figure1}, bottom panel, provides the locations of these objects.
Only 4 matches are found with the 278 early-type galaxies of Sample B:
VCC93, 209, 304 and 355.
Among the late-types galaxies, 139/191 (73\%) match an HI source. 
Thus only 52/191 (27\%) late-type VCC galaxies in Sample B do not match HI selected sources. 
Of these, 37 lie below the sensitivity threshold of ALFALFA, based on
Goldmine; the remaining 15 are not considered by us in spite of being detected, because they 
have Code$\geq$2.  \\
In any case most of these low M$_{HI}$ galaxies (32/37 or 46/52 including the code$\geq$2) are 
relatively bright and HI deficient ($Def_{HI}>0.6$).

\section{Analysis}

\subsection{Consistency with previous pointed HI measurements}
\begin{figure}
\includegraphics[width=8.5cm,angle=0]{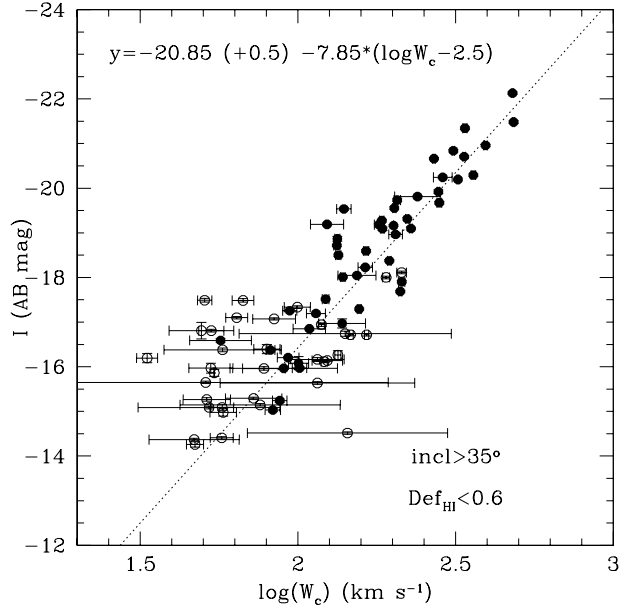}
\small{\caption{The Tully-Fisher relation as determined with ALFALFA line widths
and $i$-band asymptotic (AB) magnitudes, converted to $I_J$ and corrected for internal extinction,
limited to non-deficient galaxies with inclination
larger than 35 deg. Filled symbols represent spiral galaxies, while empty symbols refer to Irr/BCDs.
The line gives the slope of the $I$-band Tully-Fisher determination by Masters et al. (2006).} 
\label{figure4}}
\end{figure}

The comparison of the line widths determined by ALFALFA and by the pointed observations
collected in the Goldmine database is shown in Fig. \ref{figure2}.
There is a tendency ALFALFA HI line widths to be up to 50\% smaller than those of Goldmine for galaxies
with $W<100$ $\rm km~s^{-1}$. The difference decreases for increasing line widths. 
This discrepancy reflects the difference of the two methods used
to measure the line widths in the two datasets 
(i.e. average of 20\% and 50\% of the peak intensity in Goldmine vs. 50\% level of each of the two peaks
in ALFALFA).
The line measurements taken homogeneously by ALFALFA should provide a reliable
determination of the distances to individual galaxies using the Tully-Fisher relation (see next Section).\\
Fig. \ref{figure3} shows the comparison of the HI mass measurements.
There is no apparent trend with $M_{HI}$.\\

\subsection{The Tully-Fisher relation}

Using the corrected line width $W_c$ as measured by ALFALFA and the $i$-band luminosity 
derived from the total $i_{AB}$ 
magnitude, and restricting to objects with inclination larger than 35 deg,
we obtain the $i$-band Tully-Fisher relation for galaxies
in the Virgo cluster (see  Fig. \ref{figure4}). For this specific purpose it was 
necessary to consider a subsample of non-deficient galaxies, i.e. galaxies with
the HI deficiency parameter $Def_{HI}<0.6$. This is because in the ram-pressure scenario (e.g. Abadi et al. 1999)
HI ablation proceeds outside-in,
depleting first the gas that is less gravitationally bound,
so that the full width of the measured line profile underestimates the rotational velocity.
After allowing for a 0.50 mag shift for converting $i_{AB}$ (SDSS) into $I_J$ (Johnson) magnitudes (according
to the conversion table in NED), 
we apply the internal extinction corrections as in Giovanelli et al. (1997).
For $log(W_c)>2$ and excluding the Irr/BCD galaxies, the obtained relation appears consistent in slope and zero point with
the template $I_J$ Tully-Fisher determined by Masters et al. (2006).  
This indicates that the distances of the individual Virgo galaxies adopted in 
this work are not badly determined.
For $log(W_c) \leq 2$ and for Irr/BCD the dispersion of the relation makes it useless as a distance indicator.

\section{Discussion}

\subsection{The HI to stellar mass relation}

\begin{figure}
\includegraphics[width=9cm,angle=0]{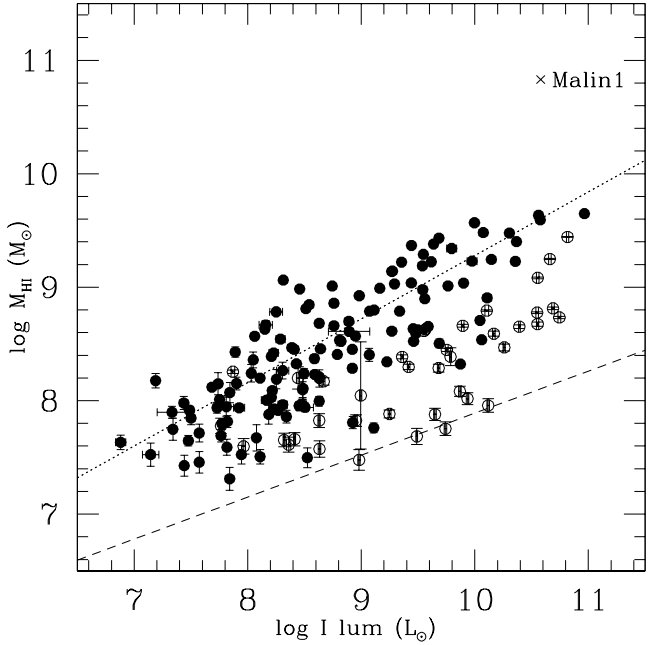}
\small{\caption{HI/optical luminosity correlation plotted separately for non-deficient (filled dots)
and for objects with $Def_{HI}\geq 0.6$ (open symbols). 
The dashed line gives the (line width dependent) limit of ALFALFA computed for an inclination of 45 deg.
The diagonal dotted lines represent the $maximum ~likelihood$ regressions fitted to the data, independently for
HI deficient and normal galaxies.}
\label{figure5}}
\end{figure}
\begin{figure}
\includegraphics[width=9cm,angle=0]{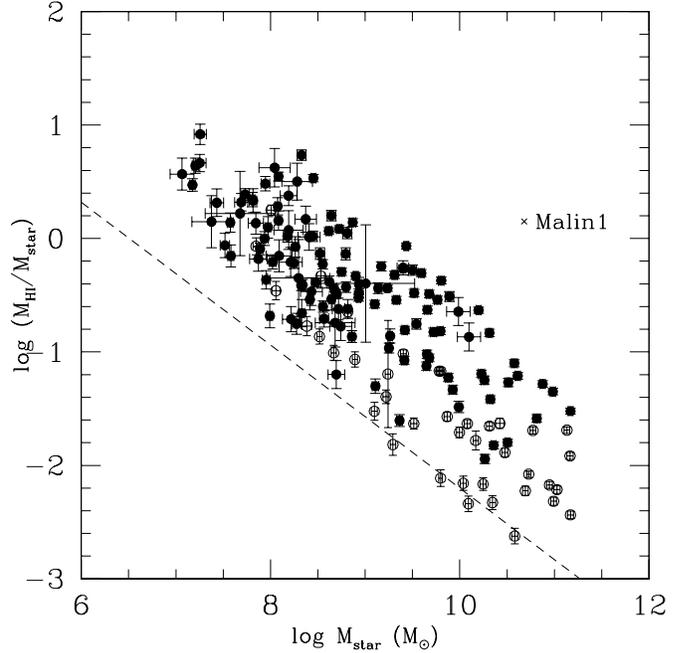}
\small{\caption{Gas to star mass ratio separately for non-deficient (filled dots)
and for objects with $Def_{HI}\geq 0.6$ (open symbols). 
The dashed line represents the limit of the HI survey computed for an inclination of 45 deg.} 
\label{figure6}}
\end{figure}

A basic feature of galaxies is the fact that   the
HI mass increase with increasing optical luminosity is flatter than the direct proportionality
(Roberts \& Haynes 1994).
Fig. \ref{figure5} shows the relation between these two quantities (given for all optically identified HI selected
sources).
Because the present analysis is carried out with galaxies belonging to
a rich cluster, the study of the dependence of HI mass on luminosity must be corrected for 
the effects of ram-pressure stripping (that causes HI deficiency: Giovanelli \& Haynes 1985) 
by excluding HI deficient objects ($Def_{HI}\geq 0.6$).
This is why galaxies with  ``normal" HI content are plotted separately from the HI-deficient objects 
in Fig. \ref{figure5}.
(We remind the reader that the definition of HI deficiency adopted here is a diameter-, not an
optical-luminosity-based relation). The line fitted to the non-deficient objects (using the "$maximum ~likelihood$" method)
shows that the slope of the relation with the optical ($i$-band)
luminosity is $<1$, with $log (M_{HI}/M_\odot)=(0.563\pm0.002) \times log(I~lum)+(3.68\pm 0.02)$. 
After transforming the optical luminosity into the stellar mass (using the relation 
given in Sect. 3.3) the meaning of the above finding is emphasized by plotting
in Fig. \ref{figure6} the ratio of the HI to stellar mass versus the stellar mass itself.
By focusing only on the non-deficient objects it is apparent that 
the most massive galaxies, the giant spirals with stellar masses in excess
of $10^{10} ~M_\odot$, have  approximately 5\% of their stellar mass in gas, intermediate
mass galaxies have 10\% gaseous mass, but in the dwarfs the gas content can exceed the mass of the stars
by a factor up to 10. 
Molecular hydrogen is not expected to change this trend.  
 Using a luminosity/metallicity dependent CO to $H_2$ conversion factor, it has been shown 
 that the molecular gas fraction decreases with increasing stellar mass and that the $MH_2/M_{HI}$ 
 ratio stays approximately
 around 15\% regardless of the morphological type (see Fig. 6 in Boselli et al. 2002).
The extrapolation of this relation to even lower stellar masses,
suggests the existence of completely gas-dominated objects, 
similar to the so called ``dark galaxies". 
One of the results of ALFALFA is however that, at least in the mass range covered
by the survey, there are practically no dark galaxies in the Virgo cluster
(Kent et al. 2007; Haynes, Giovanelli \& Kent 2007).\\
In both Fig. \ref{figure5} and \ref{figure6} we plot the position of Malin1 (taken from Pickering et al. 1997)
to show that we are sensitive to Malin1 type objects, if they existed in the Virgo cluster.\\
 As pointed out in Section 2.1, the mass limit of any flux-limited HI survey depends on the line width, 
thus on the system's mass. 
Using the virial theorem, assuming a constant mass/light ratio of 4.6 
(Gavazzi et al. 1996) and the best fit relation between $R_e$ and the stellar mass (Fig. \ref{cubeopt} and Table 1),
we derive a relation between W (the line width) and the stellar mass. Using eq. 1, 
we obtain the sensitivity of ALFALFA in HI mass units as $log (M_{HI}/M_\odot) = 4.25 +0.37*logM_{star}-0.15 $
(holding for non-deficient objects), where the last term represents the correction 
for a galaxy mean inclination of $45^\circ$. 
For less inclined galaxies the sensitivity increases, approaching $log (M_{HI}/M_\odot)  = 7.5 $.
The relation is plotted in Figs \ref{figure5} and \ref{figure6} 
representing the limiting HI mass of sources in ALFALFA as a function of the stellar mass.

\subsection{The surface brightness to stellar mass relation}

\begin{figure}
\includegraphics[width=9cm,angle=0]{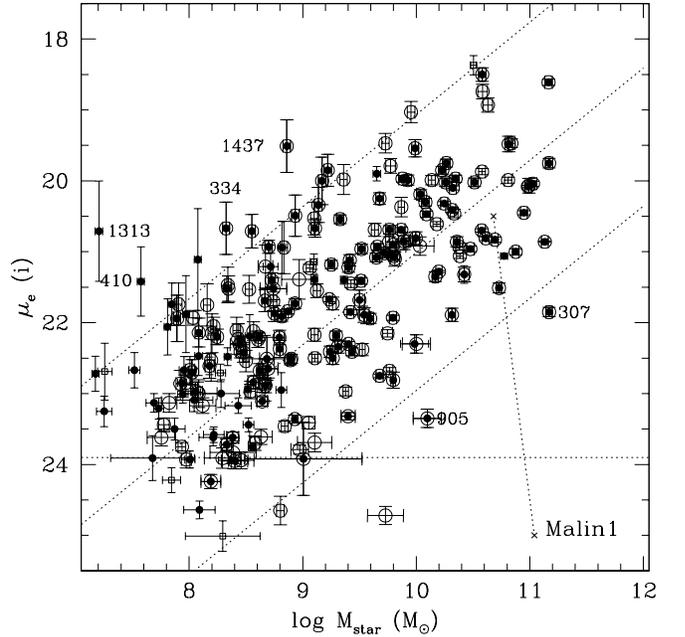}
\small{\caption{The $\mu_e$ vs. $M_{star}$ relation for HI selected late-type galaxies (filled circles),
optically selected late-type galaxies (empty circles) and HI selected early-type galaxies (squares). 
The diagonal lines represent the linear fit to the data
with $\pm 2\sigma$. Discrepant objects are labeled with their VCC names and Malin1 is given for comparison.
The horizontal line drawn at $\mu = 23.9 ~\rm mag~arcsec^{-2}$ is the $i$ band mean limiting surface
brightness of SDSS.}
\label{figure7}}
\end{figure}

One interesting issue that the ALFALFA observations of the Virgo cluster allow us to address, 
as outlined in the Introduction,  
is the question whether optically selection criteria strongly bias our knowledge of galaxies.
For example, low-surface brightness objects, particularly giant, 
low-surface brightness, gas-rich galaxies such as Malin1, are strongly undersampled in optical catalogues,  
but should be detected in a radio selected sample .
To explore this issue we exploit the ALFALFA sample in conjunction with the excellent 
(20 mag limiting magnitude and $\sim 25.3 ~\rm mag~arcsec^{-2}$ limiting surface brightness)
VCC catalogue, by comparing the optical structural parameters derived for the optically selected and for
the HI selected samples.
We first
compare in Fig. \ref{figure7} the scaling law between
the optical ($i$-band) effective surface brightness and the stellar mass  
(for the subsample of the ALFALFA or VCC galaxies that we could measure on the SDSS images,
i.e., all but 14 objects).
The data include the HI-selected sample (filled circles \& squares) and the optically-selected sample of late-type galaxies
with $m_B \leq 17$ (empty circles). The two sets of data appear to be consistent one another.
The $maximum ~likelihood$ fit to the optically selected sample is plotted together with the lines at $\pm 2\sigma$. 
The horizontal line drawn at $\mu = 23.9 ~\rm mag~arcsec^{-2}$ is the $i$ band mean limiting surface
brightness of SDSS that we have determined as 1$\sigma$ of the sky via analysis of hundreds of images.
The few measurements that lie below this line are meaningless, since they are affected by
$\sim$ 1 $\rm mag~arcsec^{-2}$ uncertainty.
There are at most two galaxies (VCC 307 and VCC 905, highlighted in Fig. \ref{figure7}) that have
slightly low surface brightness for their mass,
and four BCD galaxies (i.e. VCC 334, VCC 410, VCC 1313 and VCC 1437) 
at the opposite side of the relation, with high surface brightness 
for their mass. There are no others that mimic Malin1.
The prototype of giant low-surface brightness galaxy, Malin1, plotted for comparison, deserves
some caution. The galaxy is plotted twice (connected with a line): once at the position given by the discoverers
(Bothun et al. 1987) with $\mu_o(V)=25.5$ $~\rm mag~arcsec^{-2}$(that we transform into $\mu_o(i)=25.0 ~\rm mag~arcsec^{-2}$) at
$10^{11}~L\odot$, the other as measured on the SDSS plates. On this material the low
surface brightness extended disk is not detected, instead the galaxy appears as an amorphous enhancement of
high surface brightness ($\mu_e \sim 20.5 ~\rm mag~arcsec^{-2}$), consistent with the inner disk
detected in the HST measurement of Barth (2007).
This raises the question whether other Malin1 type objects are missing in Fig. \ref{figure7}
because their low-surface brightness disks are below the limiting surface brightness of SDSS. 
These systems would wrongly count as high surface brightness systems because we would detect only
their bulges or inner disks, possibly affecting the correlation between $\mu_e$ and the stellar mass 
or even making it spurious. 
To make sure that this is not the case 
we inspected individually all galaxies  on the SDSS and deeper plate material (Goldmine)
and we concluded that all high surface brightness objects plotted in Fig. \ref{figure7}
are genuine high surface brightness spirals and BCDs, not bulges associated with missed 
extended disks of much lower surface brightness.
The BCDs cannot be mistaken for Malin1 objects because the associated HI components
have low mass and widths not exceeding 100 $\rm km ~s^{-1}$,
opposite to the HI rich and massive Malin1. Even if they were perfectly face-on, their much bluer color would
 discriminate BCDs from possible bulges of missed Malin1 type objects. 
We conclude that
no Malin1-type galaxy is found in the Virgo cluster that we could have missed optically or
by HI selection, 
in agreement with Minchin et al. (2004), who put severe upper limits on the existence of
massive low surface brightness objects in the local field.
Moreover we find a definite trend for the $i$-band surface brightness 
to correlate with the system mass. Extremely low surface brightness galaxies
exist, but they all lie at the low mass limit of HI and optical surveys.
At the limiting sensitivity of ALFALFA at Virgo ($M_{HI} \sim 10^{7.5 \sim 8} ~M_\odot$)
there are at most 10/185 HI sources that could possibly be identified with galaxies of lower
surface brightness than the SDSS limit ($\sim 23.9 \rm ~mag ~arcsec^{-2}$).  \\
\begin{figure*}
\centering
\includegraphics[width=13cm,angle=0]{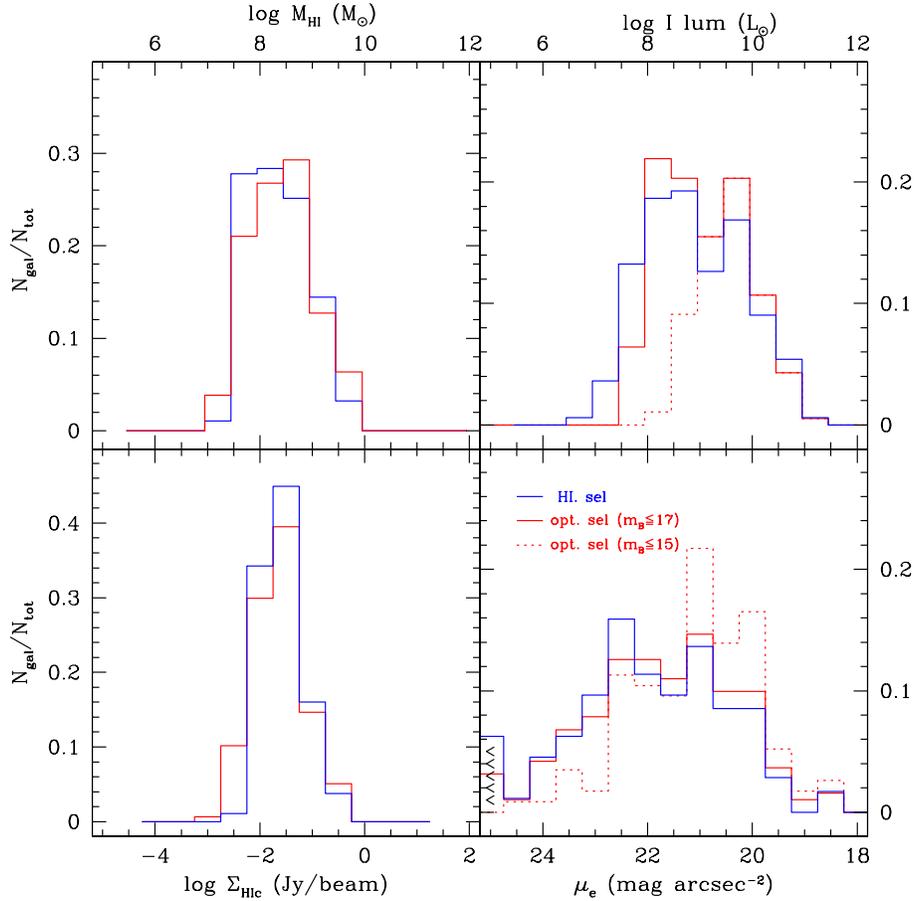}
\small{\caption{Histograms of luminosities (top) and surface brightness (bottom) for the
HI-selected (left) and optically-selected (right) samples.
The distributions refer to the HI-selected sample (blue) and to 
the optically-selected sample limited to $m_B \leq 17$ (solid red)  and to
$m_B \leq 15$ (dotted red). Note that the bins at $\mu_e=25~\rm ~mag ~arcsec^{-2}$ represent upper limits
(10 HI selected and 6 optically selected objects too faint to be measured in $i$-band).}
\label{figure8}}
\end{figure*}

\subsection{The distribution of mass and surface brightness}

Fig. \ref{figure8} contains histograms of the distribution of surface brightness and luminosity,
both HI and optical, as derived for optical (red) and HI selected  samples (blue).
It may be surprising that there appears to be very little difference between the red and 
the blue histograms. 
There is indeed some excess of low surface brightness objects
in the HI selected sample, represented by the 10 very faint galaxies that could not be measured 
on the SDSS material and are plotted
as upper limits in the $\rm 25~mag~arcsec^{-2}$ bin. (Note that the 10 tidally-disrupted systems 
were not counted among the low-surface brightness objects.) 
However these sources amount to less than 5 \% of the total number.
Similarly in the optically-selected sample there are 4 VCC galaxies 
that could not be measured on the SDSS material.
The difference is surprisingly small when we compare the HI-selected sample with 
the $m_B \leq 17$ optically-selected VCC sub-sample. To show that this agreement is not at all obvious 
we compute the distribution also for a shallower subsample of the VCC, limited to $m_B \leq 15$ (red dotted
histogram). 
This time the optical (luminosity and surface brightness) distributions differ significantly
from the HI-selected distributions, owing to the existence of the surface brightness vs.
luminosity correlation that was investigated in Section 5.2 (see Fig. \ref{figure7}).
Virgo appears just at the right distance from us to make a survey with the sensitivity of ALFALFA 
to match almost exactly the VCC galaxies of late-type with $m_B \leq 17$.\\

\subsection{Other optical structural parameters}

\begin{table*}

  \caption{$Maximum ~likelihood$ regressions}
  \label{Tabtrans}
  {\scriptsize
  \[
  \begin{array}{lllcrr}
  \hline
  \noalign{\smallskip}
y & x & Condition & Figure & slope & intercept\\
\hline
\log M_{HI}  &  \log I~lum  &Def_{HI}<0.6   &   \ref{figure5}    &  0.563 \pm 0.002 & 3.68 \pm 0.02 \\
\mu_e & \log M_{star}      & opt             &   \ref{figure7}   &  -1.300\pm 0.008 & 34.01\pm 0.08\\
\log M_{star} &  g-i  & opt &  \ref{cubeopt}	&  4.83 \pm 0.08& 5.77\pm 0.07 \\
\log M_{star} &  g-i  & HI & \ref{cubeopt}	&  4.42 \pm 0.07 & 6.34 \pm 0.05\\
\log M_{star}  &  \log R_e & opt     &    \ref{cubeopt}	&  3.97\pm0.07 & 7.48\pm 0.04\\
\log M_{star}  &  \log R_e & HI     &    \ref{cubeopt}	&  3.76\pm0.06& 7.54\pm 0.04 \\
\log M_{star}&  \mu_e   & opt  &     \ref{cubeopt}	& -0.769 \pm 0.005 & 26.16\pm 0.10\\
\log M_{star}&  \mu_e   & HI   &     \ref{cubeopt}	& -0.825 \pm 0.006& 27.37\pm 0.13\\
 g-i &  \mu_e & opt    &	       \ref{cubeopt}	& -0.155\pm 0.003 & 4.10\pm 0.07\\
 g-i &  \mu_e & HI    &	       \ref{cubeopt}	&  -0.197\pm 0.004& 4.94\pm 0.08\\
   \noalign{\smallskip}
     \hline
  \noalign{\smallskip}

  \noalign{\smallskip}
  \hline
\end{array}
  \]
  }

\end{table*}

\begin{figure*}
\includegraphics[width=18cm,angle=0]{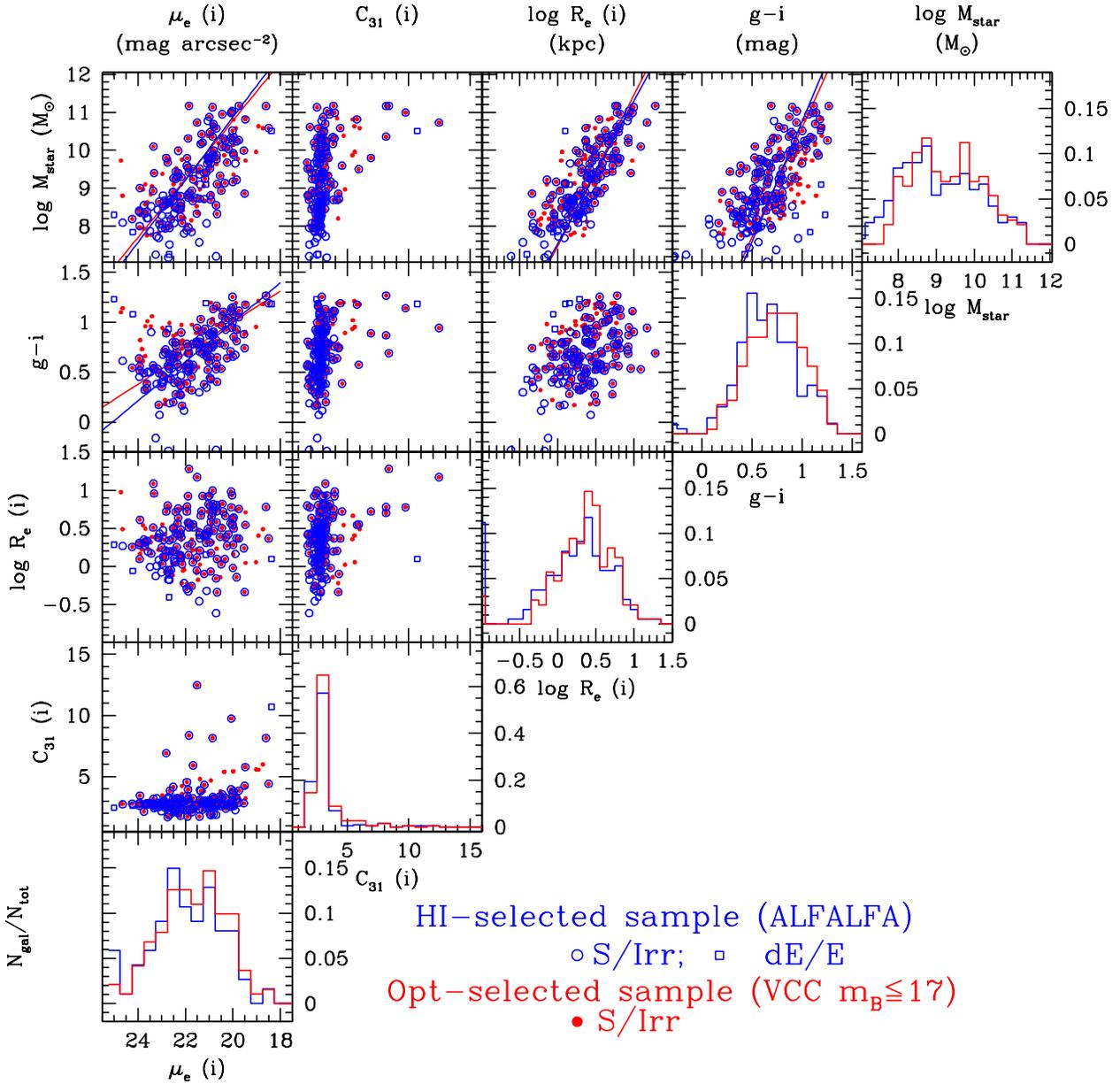}
\small{\caption{Optical structural parameters for the HI- (blue) and
optically-selected $m_B \leq 17$ late-type galaxies (red). See text for explanation of parameters. Lines represent the $maximum ~likelihood$ regressions fit to the data (see Table 1).}
\label{cubeopt}}
\end{figure*}

Inspired by the analysis carried out by Cortese et al. (2007) who studied the relations between
various structural parameters in their Fig. 6, we derive
additional optical ($i$-band) structural parameters
for the HI selected objects and for the $m_B \leq 17$ optically selected late-type VCC sub-sample.
These are the galaxy ``scale'' (stellar mass, $R_e$, luminosity or combinations) and 
``form''($\mu_e$, $g-i$, $C_{31}$) (following the terminology of Whitmore 1984).
They are plotted in Fig. \ref{cubeopt} for 
sources identified with galaxies selected in the blind ALFALFA HI survey (blue)
and for the optically selected late-type systems from the VCC catalogue, limited to $m_B \leq 17$ (red).
The $maximum ~likelihood$ regressions fitted to the data are plotted in Fig. \ref{cubeopt} and reported in Table 1.\\
It is evident from Fig. \ref{cubeopt} that the scaling relations well known 
for optically selected galaxies (e.g. color vs. mass, color vs. surface brightness, $C_{31}$
 vs. mass; Gavazzi et al. 1996, Scodeggio et al. 2002),
including the Kormendy relation ($R_e$ vs. $\mu_e$), also hold true for HI selected objects.
We emphasize that this result applies to galaxy members of a rich cluster such as Virgo. 
For example the $\mu_e$ vs. $M_{star}$ relation clearly shows that low surface brightness
galaxies exist (and are easy to miss), but they lie mostly at the low end of the luminosity function,
as discussed above. Notice in the color vs. stellar mass and color vs. surface brightness panels 
that the deviant points (faint-red objects) are generally dE 
that have structural and photometric properties intermediate between quiescent and star forming dwarfs
(Boselli et al. 2008) \\
Furthermore the frequency distributions of the ``scale'' and ``form'' parameters
are indistinguishable in HI and optically selected surveys. For example the frequency 
of pure disks, that have $C_{31}<3$ exceeds the frequency of bulge+disk systems ($C_{31}>>3$) in both 
the HI- and optically-selected samples. But galaxies with $C_{31}>>3$ galaxies exist in both
samples. 
We conclude that the scaling laws that exist among the optical ($i$-band) structural parameters 
of disk galaxies do not differ significantly in HI and optically selected samples, although we should
emphasize again that this result is limited to a rich nearby cluster.\\ 

\section{Conclusions}

By surveying the Virgo cluster, ALFALFA has provided definite
evidence that: 
i) HI inhabits galaxies that are structurally similar to ordinary late-type galaxies;  
ii) their HI content can be predicted from their optical luminosity;
iii) low surface brightness galaxies have low optical luminosity and contain small
quantities of neutral hydrogen; 
iv) Malin1 type galaxies are comfortably rare objects (less than 0.5 \%);
v) there are no ``dark-galaxies" with HI masses $M_{HI} \geq 10^{7.5 \sim 8} ~M_\odot$ (see Kent et al. 2007 and
Haynes, Giovanelli and Kent 2007);  
vi) less than 1\% of early-type galaxies 
contain neutral hydrogen with $M_{HI} \geq 10^{7.5 \sim 8} ~M_\odot$ (di Serego Alighieri et al. 2007).\\
Once ALFALFA covers a substantial fraction of its final area, including the Virgo cluster
in its full extent, a determination of the HI mass function will be obtained with unprecedented significance.
 We expect that the observed HI mass function of the cluster can be predicted with sufficient accuracy
from the optical luminosity function of the late-type galaxies alone, accounting for 
the HI vs. optical luminosity relation discussed in Section 5.1 (Fig. 5),
computed separately for normal and HI deficient objects (Kent et al. in preparation).
If this is the case we expect that the faint-end slope will not differ significantly from that of the 
optical luminosity function of late-type galaxies, i.e. flatter than the halo mass function
predicted theoretically from the CDM cosmology.
\acknowledgements 
We thank L. Cortese and C. Bonfanti for precious hints and useful discussions.
This research has made use of the Goldmine database. \\
 Funding for the Sloan Digital Sky Survey (SDSS) and SDSS-II has been provided by the 
 Alfred P. Sloan Foundation, the Participating Institutions, the National Science Foundation, 
 the U.S. Department of Energy, the National Aeronautics and Space Administration, 
 the Japanese Monbukagakusho, and 
 the Max Planck Society, and the Higher Education Funding Council for England. 
 The SDSS Web site is http://www.sdss.org/.
 The SDSS is managed by the Astrophysical Research Consortium (ARC) for the Participating Institutions. 
 The Participating Institutions are the American Museum of Natural History, Astrophysical Institute Potsdam, 
 University of Basel, University of Cambridge, Case Western Reserve University, The University of Chicago, 
 Drexel University, Fermilab, the Institute for Advanced Study, the Japan Participation Group, 
 The Johns Hopkins University, the Joint Institute for Nuclear Astrophysics, the Kavli Institute for 
 Particle Astrophysics and Cosmology, the Korean Scientist Group, the Chinese Academy of Sciences (LAMOST), 
 Los Alamos National Laboratory, the Max-Planck-Institute for Astronomy (MPIA), the Max-Planck-Institute 
 for Astrophysics (MPA), New Mexico State University, Ohio State University, University of Pittsburgh, 
 University of Portsmouth, Princeton University, the United States Naval Observatory, and the University 
 of Washington.


\begin{thebibliography}{}

\bibitem[]{} Abadi, M. G., Moore, B., Bower, R. G., 1999, MNRAS, 308, 947
\bibitem[]{} Adelman, J., et Gal. 2007, Ops (in press)
\bibitem[]{} Build, R., et ASL. 2006, MENS, 371, 1617
\bibitem[]{} Barth, A.J., 2007, AJ, 133, 1085
\bibitem[]{} Bell, E. F. et al., 2003, ApJ, 149, 289
\bibitem[]{} Binggeli, B., Sandage, A., Tammann, G., 1985, AJ, 90, 1681
\bibitem[]{} Binggeli, B., Popescu, C., Tammann, G., 1993, A\&AS, 98, 275
\bibitem[]{} B\"ohringer H., et al. 1994, Nat, 368, 828
\bibitem[]{} Blanton, M.R., Hogg, D.W., Bahcall, N.A., et al., 2003, ApJ, 592, 819
\bibitem[]{} Boselli, A. \& Gavazzi, G., 2006, PASP, 118, 517
\bibitem[]{} Boselli, A., Gavazzi, G., Donas, J. \& Scodeggio, M., 2001, AJ, 121, 753
\bibitem[]{} Boselli, A., Lequeux, J. \& Gavazzi, G., 2002, A\&A, 384, 33
\bibitem[]{} Boselli, A., Boissier, S., Cortese, L., Gavazzi, G., 2008, ApJ, in press
\bibitem[]{} Briggs, F. H., 1997, ApJ, 484, L29
\bibitem[]{} Cortese, L., et al., 2007, MNRAS, in press
\bibitem[]{} de Vaucouleurs, G., 1977, in ``Evolution of Galaxies and Stellar
 Populations'', eds. R. Larson \& B. Tinsley (New Haven: Yale University Observatory), 43
\bibitem[]{} Disney, M., 1976, Nature, 263, 573
\bibitem[]{} Disney, M., \& Phillipps, S., 1987, Nature, 329, 203
\bibitem[]{} Gavazzi, G., Pierini, D. \& Boselli, A., 1996, A\&A, 312, 397  
\bibitem[]{} Gavazzi, G., Boselli, A., Scodeggio, M., Pierini, D., Belsole, E., 1999, MNRAS, 304, 595
\bibitem[]{} Gavazzi, G., Zibetti, S., Boselli, A., Franzetti, P., Scodeggio, M., Martocchi, S., 2001, A\&A, 372, 29 
\bibitem[]{} Gavazzi, G., Boselli, A., Donati, A., Franzetti, P. \& Scodeggio, M., 2003, A\&A, 400, 451
\bibitem[]{} Gavazzi, G., Boselli, A., van Driel, W., O'Neil, K., 2005, A\&A, 429, 439
\bibitem[]{} Giovanelli, R., Haynes, M., 1985, ApJ, 292, 404
\bibitem[]{} Giovanelli, R., Haynes, M., Kent, B., et al., 2005, AJ, 130, 2598
\bibitem[]{} Giovanelli, R., Haynes, M., Kent, B., et al., 2007, AJ, 133, 2569	
\bibitem[]{} Haynes, M., \& Giovanelli, R., 1984, AJ, 89, 758
\bibitem[]{} Haynes, M.P., Giovanelli, R., and Kent, B.R., 2007, ApJ, 665, L19
\bibitem[]{} Hayward, C. C., Irwin, J. A., Bregman, J. N., 2005, ApJ, 635, 827
\bibitem[]{} Impey, C.,  \& Bothun, G., 1989, ApJ., 341, 891
\bibitem[]{} Jenkins, A., Frenk, C.S., \& White, S., et. al., 2001, MNRAS, 321, 372 
\bibitem[]{} Kent, B., Giovanelli, R, Haynes, M., et. al., 2007, ApJ, 665, L15
\bibitem[]{} Kormendy, J., 1985, ApJ, 295, 73
\bibitem[]{} Kovac, K., P.h.D. Thesis, 2007, University of Groningen. 
\bibitem[]{} Masters, K. L., Giovanelli, R., Haynes, M.,  2004, Apj, 607,115, 2004
\bibitem[]{} Masters, K., et al., 2006, ApJ, 653, 861
\bibitem[]{} Meyer M. J., et al., 2004, MNRAS, 350, 1195
\bibitem[]{} Minchin, R.F., et al., 2003, MNRAS, 346, 787
\bibitem[]{} Minchin, R.F., et al., 2004, MNRAS, 355, 1303
\bibitem[]{} Pickering, T.E., Impey, C.D., van Gorkom, J.H., \& Bothun, G.D., 1997, AJ, 114, 1858
\bibitem[]{} Press, W. H., \& Schechter, P., 1974, ApJ., 187, 425
\bibitem[]{} Roberts, M.S., \& Haynes, M.P., 1994, ARAA, 32, 115
\bibitem[]{} Rosenberg, J. L., Schneider, S.E., 2002, ApJ, 567, 247
\bibitem[]{} Sabatini, S., Davies, J., Scaramella, R., Smith, R., Baes, M., Linder, S.~M., Roberts, S., \& 
             Testa, V., 2003, MNRAS, 341, 981 
\bibitem[]{} Scodeggio, M., Gavazzi, G., Franzetti, P., Boselli, A., Zibetti, S., Pierini, D., 2002, A\&A, 384, 812
\bibitem[]{} di Serego Alighieri, S. et al., 2007, A\&A (in press)
\bibitem[]{} Springob, C. M., Giovanelli, R., Haynes, M.,  2005, ApJ, 621, 215
\bibitem[]{} Warren, B.E., Jerjen, H., Koribalski, B.S.,  2006, AJ, 131, 2056
\bibitem[]{} Whitmore, B.C., 1984, ApJ, 278, 61
\bibitem[]{} Zwaan, M.~A., Briggs,  F.~H., \& Sprayberry, D., 2001, MNRAS, 327, 1249 
\bibitem[]{} Zwaan, M.A., et. al., 2003, AJ, 125, 2842
\bibitem[]{} Zwaan, M.A., et. al., 2005, MNRAS, 359, L30

\end{thebibliography}
\end{document}